\documentclass[]{emulateapj}
\usepackage[colorlinks = true, linkcolor = blue, urlcolor  = blue, citecolor = blue, anchorcolor = blue]{hyperref}
\usepackage{hyperref}

\newcommand\fnurl[2]{\href{#2}{#1}\footnote{\url{#2}}}

\shorttitle{Global Oscillation Pattern in Succeeding Solar Flares}
\shortauthors{N. Gyenge \& R. Erd\'elyi}

\begin{document}

\title{Periodic Recurrence Patterns In X-ray Solar Flare Appearances}

\author{N. Gyenge\altaffilmark{1,2,3*} \& R. Erd\'elyi\altaffilmark{1,3}}
\thanks{\altaffilmark{*}e-mail: n.g.gyenge@sheffield.ac.uk}
\affil{
\altaffilmark{1}Solar Physics and Space Plasmas Research Centre (SP2RC), School of Mathematics and Statistics,\\ University of Sheffield Hounsfield Road, Hicks Building, Sheffield S3 7RH, UK\\
\altaffilmark{2}Debrecen Heliophysical Observatory (DHO), Research Centre for Astronomy and Earth Sciences, \\Hungarian Academy of Sciences, Debrecen, P.O.Box 30, H-4010, Hungary\\
\altaffilmark{3}Dept. of Astronomy, E\"otv\"os L\'orand University, P\'azm\'any P\'eter s\'et\'any 1/A, \\Budapest, H-1117,Hungary\\
}

\begin{abstract}

 The temporal recurrence of micro-flare events is studied in a time interval before and after of  major solar flares. Our sample is based on the x-ray flare observations by the Geostationary Operational Environmental Satellite (GOES) and Reuven Ramaty High Energy Solar Spectroscopic Imager (RHESSI). The analysed data contains 1330/301 M- and X-class GOES/RHESSI energetic solar flares and 4062/4119 GOES/RHESSI micro-flares covering the period elapsed since 2002. The temporal analysis of recurrence, by Fast Fourier Transform (FFT), of the micro-flares shows multiple significant periods. Based on the GOES and RHESSI data, the temporal analysis also demonstrates that multiple periods manifest simultaneously in both statistical samples without any significant shift over time. In the GOES sample, the detected significant periods are: $11.33$, $5.61$, $3.75$, $2.80$ and $2.24$ minutes. The RHESSI data shows similar significant periods at $8.54$, $5.28$, $3.66$, $2.88$ and $2.19$ minutes. The periods are interpreted as signatures of standing oscillations, with the longest period ($P_{1}$) being the fundamental and others as higher harmonic modes. The period ratio of the fundamental and higher harmonics ($P_{1}/P_{N}$) is also analysed. The standing modes may be signatures of global oscillations of the entire solar atmosphere encompassing magnetised plasma from photosphere to corona in active regions.
 
 \end{abstract}

\section{Solar Atmospheric Oscillations}

The detailed understanding of the nature of the solar atmospheric intensity oscillations is a long-standing challenge. The intensity oscillations could provide vital information about the properties of the solar atmosphere (e.g. geometric structure, magnetic structure, density structure, ionisation degree) by using solar magneto-seismology tools \citep{roberts1984coronal, banerjee2007present, erdelyi2008hinode, 2009SSRv..149....3A, verth2010magnetoseismology, 2015SSRv..190..103J}. Numerous studies reported oscillations in the solar atmosphere, using high-resolution observations \citep{de2009longitudinal, ruderman2009transverse, banerjee2011propagating, wang2011standing, mathioudakis2013alfven}. The observed periods of intensity oscillations range from several minutes to several hours \citep{auchere2014long}. Various oscillation patterns with periods of few dozens of minutes  are also found in polar plumes and polar coronal holes observations \citep{deforest1998observation}. \cite{bocchialini2011oscillatory} studied intensity and Doppler velocity oscillations and reported periods from several up to 80 minutes in filament and prominence observations. \cite{tian2008long} investigated solar bright points and reported oscillations between periods of 8 to 64 minutes. Often magnetohydrodynamic (MHD) waves in the solar corona are accounted for the observed intensity oscillations in the range of 2-33 minutes \citep{aschwanden2002transverse}. Hence, long-period oscillations in the solar atmosphere are not unprecedented. 

Smaller local features like solar flares are also able to produce a periodic behaviour and studying their oscillatory patterns became a well-studied subject as well \citep{2018SSRv..214...45M}. In general, the observed periodic features in the wavelet power or the Fourier spectrum of the soft x-ray emissions are called quasi-periodic pulsations (QPP). Based on the observations of the GOES satellite, \cite{2012ApJ...749L..16D} performed a case study that reported QPP signatures in the emission of an X-class solar flare. \cite{2015SoPh..290.3625S} confirmed these results by analysing a larger statistical sample. They demonstrated that 28 events out 35 X-class flares also show QPP signatures. \cite{2011A&A...525A.112R} used RHESSI observations to show that periodicities in the range of $2.5 - 5.0$ minutes become shorter the closer the observations are to a major energetic flare. \cite{2009A&A...505..791S} found similar QPP periodicities and proposed that the source of the oscillations could be triggered by 3-minute slow magnetoacoustic waves. \cite{2015A&A...577A..43S} found continuous energy amplification of 3-minute waves in sunspot umbra before a solar flare.

The observed oscillations in coronal loops may indicate that standing slow modes are likely triggered by micro-flares, which are produced by impulsive heating \citep{mendoza2002coronal, taroyan2005footpoint, erdelyi2008hinode}. On the other hand, the micro-flares themselves before a major, energetic flare are usually called precursors \citep{charikov2000x} and their temporal distribution may be linked to other types of periodic variations of x-ray flux. Namely, the majority of the hard x-ray flares are preceded by precursors, usually a few dozens of minutes before the major solar flare \citep{tappin1991all}. The information obtained by the observations of these flare precursors can be applied for acquiring the spatio-temporal properties of the local magnetic the reorganisation process and may also reveal diagnostic information about the nature of the destabilisation of the active region. 

In this work, we study the temporal distribution of solar flare (mostly micro-flare) recurrences before and after an energetic eruption. The energetic eruption, here, refers to X- and M-type of solar flares. Micro-flare occurrence before or after a major flare will be referred to as pre- or post-flare activity. Furthermore, for the sake of simplicity, both types of micro-flares, pre- or post-flares, will be jointly referred to as minor-flares.

\section{Methodology}

Two x-ray flare databases are employed in our study. Firstly, the major- and minor-flares are provided by the GOES satellite. The GOES catalogue contains information about the basic properties of the solar flares, such as the onset time, the position, the magnitude of the events and the identification of the associated active region. The flare catalogue is available at \fnurl{NGDC/NOAA}{ftp://ftp.ngdc.noaa.gov/STP/space-weather/solar-data/solar-features/solar-flares/x-rays/goes/}. Although the main focus is on the statistical population of the GOES eruptions, a control sample is also used which is based on the \fnurl{flare list}{http://hesperia.gsfc.nasa.gov/hessidata/dbase/hessi_flare_list.txt} by the RHESSI satellite \citep{lin2002reuven}. The RHESSI flare list contains data about the onset time of the flare, duration of the event, peak intensity, photon count and energy channel of the maximal energy. The position of the solar flares are calculated by 128x128 back-projection maps using 16-arcsecond pixels \citep{hurford2003rhessi}. The spatial resolutions may seem to be somewhat inaccurate, the position data are sufficiently accurate for locating the active region. The GOES and RHESSI catalogues contain 25691 and 121430 solar flare events each, respectively,  for the analysed time period between 2002 and 2017. The discrepancy between the number of observed solar flares may lie in the sensitivity of satellite detectors and/or the significance threshold of the signal processing. Although, the definition of identification of solar flares is similar for both catalogues, the two satellites observes in slightly different wavelength range, which may also cause further discrepancies.

For identifying major flares for both the GOES and RHESSI samples the following criteria are introduced. In the GOES statistical sample, only M- and X-class flares are selected as a major flares. However, the RHESSI data does not contain flare classification. Therefore, the RHESSI major flare candidates must be associated by the GOES  counterpart records for obtaining the flare classification information for each solar flare. Only simultaneously observed events are considered, i.e. the RHESSI and GOES solar flares must be relatively close in space and time. The actual information of RHESSI major flares are taken from the RHESSI flare database. The GOES flare counterpart event only assists in filtering the magnitude of the RHESSI observations. It is also required that no additional flare occurs with a larger peak flux in the same sunspot group within a 6-hour interval before and after the candidate major flare. A major flare candidate without an associated active region is not considered for selection for further analysis.

\begin{table}
	\centering
		\caption{Number of Major Events Before and After Filtering}
			\label{table0}
		\begin{tabular}{llrr}
			Source & Type & Before Filtering & After Filtering  \\
			\\
			GOES  & M-class  & 1340 & 1219\\
			GOES & X-class  & 115 & 111\\
			\hline 
			Total &   & 1455  & 1330 \\
			\\
			RHESSI & M-class & 593 & 290  \\
			RHESSI & X-class & 56 & 11  \\
			\hline 
			Total &  &  649 &  301\\	
			\\
		\end{tabular}
	\label{table0}
\end{table}

\begin{table}
	\vspace{-2em}
	\centering
	\caption{Number of Minor Events Before and After Filtering}

		\begin{tabular}{llrr}
			Source & Type & M-class Major &  \\
						 &         & Before Filtering & After Filtering \\

			GOES & Pre-flare & 2270  & 2102 \\
			GOES & Post-flare & 1807 & 1675 \\
			\hline 
			Total &  & 4077 &  3777 \\
			\\
			RHESSI & Pre-flare & 4273 & 1851 \\
			RHESSI & Post-flare & 4043 &  2067 \\
			\hline 
			Total &  & 8316 &   3918 \\		
			\\
			Source & Type & X-class Major &  \\
						 &         & Before Filtering & After Filtering \\

			GOES & Pre-flare & 191  & 188 \\
			GOES & Post-flare & 100 & 97 \\
			\hline 
			Total &  & 291 &  285 \\
			\\
			RHESSI & Pre-flare & 534 &  109\\
			RHESSI & Post-flare & 441 &  92 \\
			\hline 
			Total &  & 975 &  201  \\	
			\\
			Source & Type & All Major &  \\
						 &         & Before Filtering & After Filtering \\

			GOES & Pre-flare & 2461  &  2290\\
			GOES & Post-flare & 1907 & 1772 \\
			\hline 
			Total &  & 4368 & 4062   \\
			\\
			RHESSI & Pre-flare & 4807 &  1960\\
			RHESSI & Post-flare & 4484 &   2159\\
			\hline 
			Total &  & 9291 &  4119  \\		
		\end{tabular}

	\label{table1}
\end{table}

\begin{figure*}[t]
	\centering
	\includegraphics[width=\textwidth]{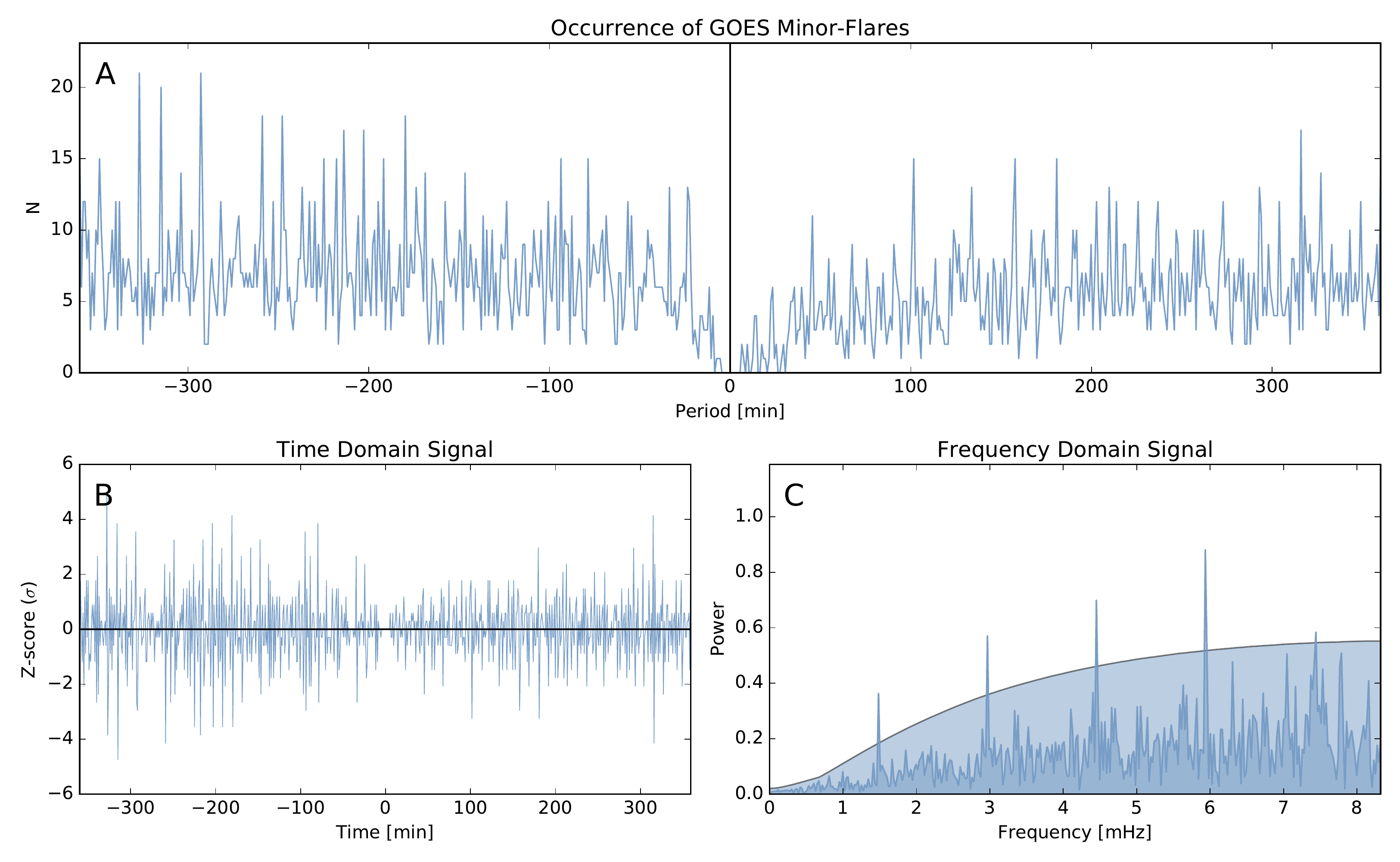}
	\caption{Panel $A$ demonstrates the temporal variation of the pre- and post-flare occurrences before and after the major M- or X-class flare based on the GOES flare data. Panel $B$ represents the time domain signal. This signal is transformed by applying the first difference on the original signal. Panel $C$ represents the power spectrum of the FFT. Under the blue area the peaks are not considered as significant frequencies. These peaks are below the $3 \sigma$ significant threshold.}
	\label{fig1}
\end{figure*}	

Now, the final GOES sample obtained by applying the above criteria contains $1330$ M- and X-class events (including $1219$ M-class and $111$ X-class event) between $2002$ and $2017$. The final RHESSI statistical population consists of $301$ major flares, $290$ M- and $11$ X-class events between the same period as the GOES sample. Table \ref{table0} shows the number of major flare events before and after filtering. The total of the RHESSI sample may seem somewhat low compered to the sample size of the GOES sample. Due to the orbital properties of the RHESSI satellite half of the solar eruptions cannot be observed. A significant portion of the data is also lost because of the missing active region identification, the missing GOES counterpart flare association, the influence of the orbit of the RHESSI satellite and other interferences, such as the South Atlantic Anomaly \citep{christe2008rhessi}.

\begin{figure*}[t]
	\centering
	\includegraphics[width=\textwidth]{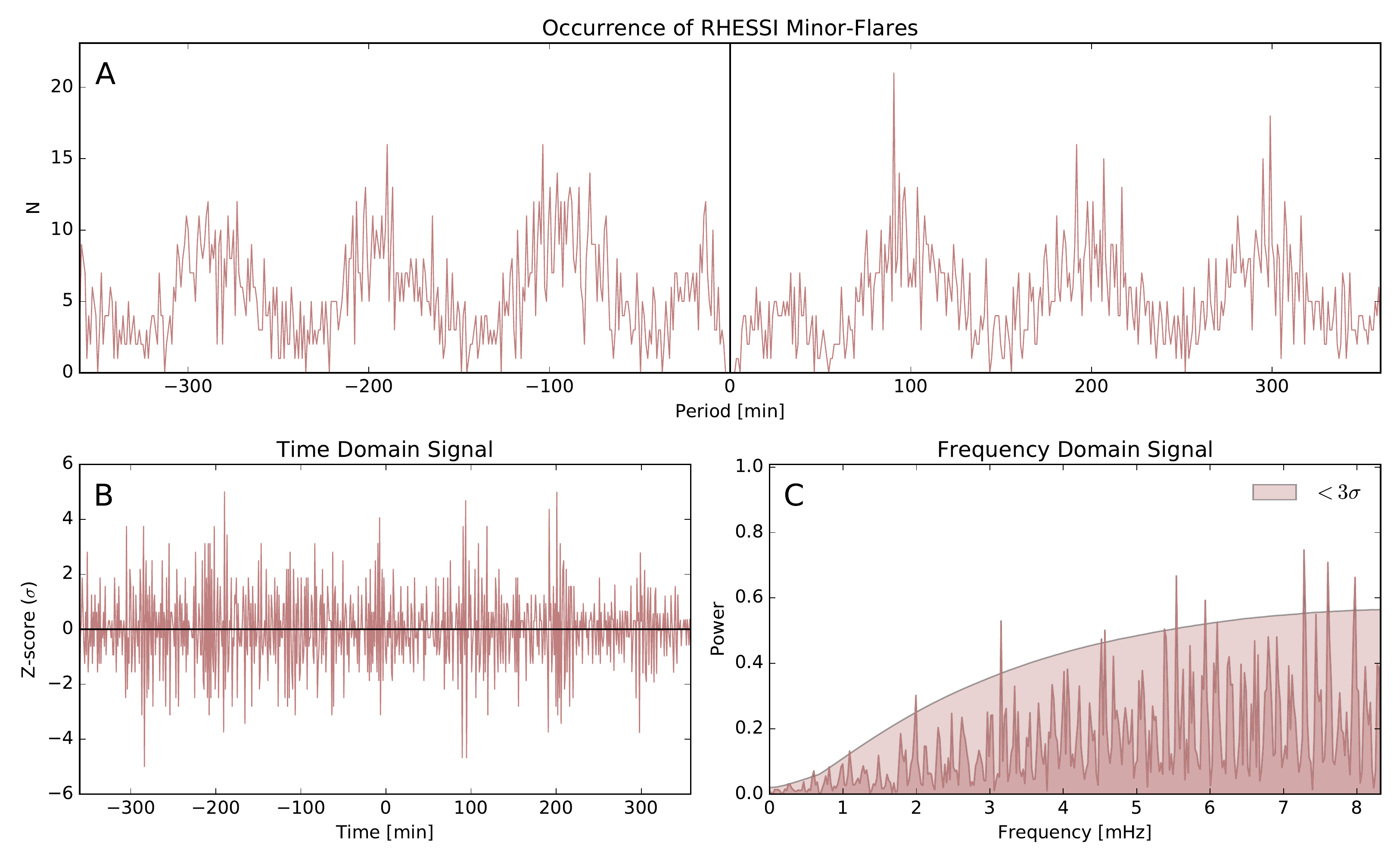}
	\caption{Panel $A$ shows the time variation of the minor-flare occurrences before and after the major flare based on the RHESSI statistical sample. Panel $B$ demonstrates the transformed data by employing first difference and seasonal difference techniques. The Panel $C$ is the power spectrum of the FFT. Under the red area the peaks are not considered as significant frequencies. These peaks are below the $3 \sigma$ significance threshold.}
	\label{fig2}
\end{figure*}	

Criteria for filtering the minor-flares are also applied. The RHESSI satellite's orbit allows one hour observation time then 40 minutes observation black out caused by the satellite spending time in orbital eclipse. Furthermore, when the satellite passes through the intermediate southern latitudes the detector counts are influenced by the South Atlantic Anomaly. Therefore, we omitted solar flares that occurred crossing the South Atlantic Anomaly and data gaps. More specifically, the RHESSI eruptions indicated by the flag ED, EE, ES, DE ,DS, DG, DE, GS, NS SD, SE, SS and PS are omitted. Fortunately, the GOES flare catalogue does not suffer periodic data gaps by the satellite orbit as the RHESSI observations. When a minor-flare occurs in the same sunspot group within at most 6 hours before or after the major flare, the eruption is automatically considered as a minor-flare of a major flare. However, the minor flares must be less energetic than the associated major flare. In the GOES sample, the flare classification of the minor flare candidate must be smaller than the classification of the major event. In case of the RHESSI data, the highest energy band in which the minor event is observed must be smaller than the highest energy band of the major solar flare. At this stage, direct physical causality between the major and the minor-flare cannot be assumed, however, both events could be the consequence of the reorganisation of local magnetic field. After applying our introduced filtering criteria, the total number of GOES minor-flare events is $4062$, containing $2290$ pre-flares and $1772$ post-flares. Meanwhile, the RHESSI minor-flare population is composed by $4119$, including $1960$ pre-flares and $2159$ post-flares. The Table \ref{table1} demonstrates the number of minor flare events before and after applying the filtering criteria.

Let us define the reference time as the moment of the major flare eruption for each active region. Next, the elapsed time between the eruption of minor-flare $(t_{i})$ and their major flare $(t^{*})$ is calculated for each major flares in every active region separately. Let us now introduce, 

\begin{equation}
	A_{n}= \{ (t_{1} - t^{*}) ,(t_{2} - t^{*}),(t_{3} - t^{*}), ...,(t_{i} - t^{*}) \},
	\label{eq1}
\end{equation}

\noindent
where $A_n$ contains the time differences between the major and minor solar flares in each active regions. The actual value of $(t_{i} - t^{*})$ must be between $-360$ minutes and $360$ minutes, however, $0\not\in A$. The domain $[-360, 0[$ represents the pre-flares up to 6 hours prior to the major flare and $]0, 360]$ stands for the post-flares, up to 6 hours after the onset of the major event. The subscript $n$ represents the number of major flares. The total number of elements for a given $A_{n}$ equals the total number of pre- and post-flares. On average, each GOES major flares are surrounded by $3$ GOES minor-flares and each RHESSI major flare has $13$ RHESSI minor-flares. Hence, case studies based on a single active region cannot be performed. To increase the number of events we merge all major flares $(n)$ in all active region into set $x$. Therefore, the total statistical sample size is now defined by the expression,

\begin{equation}
	x = \bigcup_{i=1}^{n} A_{i}.
	\label{eq2}
\end{equation}

The frequency distribution $F(x)$ is calculated in one-minute bins of the 6-hour period before and after the main flare, therefore the number of bins is 720. In each bin, we determine the total number of the minor flares. One bin still contains approximately half a dozen minor-flares, more specifically in the GOES statistics each bin contains 5.6 solar flares on average. In case of the RHESSI population, 5.7 solar flares are present per bin on average. The frequency distribution $F(x)$ is normalised by the following definition (also referred to as Z-Scores):

\begin{equation}
	Z(x_{i}) = \frac {F(x_{i}) - \overline{F(x)}}  {\sigma(F(x))},
	\label{eq3}
\end{equation}

\noindent
where, $\overline{F(x)}$ represents the mean of the frequency distribution $F(x)$ and $\sigma(F(x))$ is the standard deviation. The mean $\overline{F(x)}=5.64$ and standard deviation $\sigma(F(x)) = 3.37$ for the GOES statistical population. The RHESSI sample shows the mean $\overline{F(x)}=5.72$ and the standard deviation $\sigma(F(x)) = 3.20$. Finally, anomalies or outliers were identified and excluded from the further statistics, therefore we omitted peaks greater than $|Z(x_{i})| > 5 \sigma$ threshold.

\begin{figure*}
	\centering
	\includegraphics[width=\textwidth]{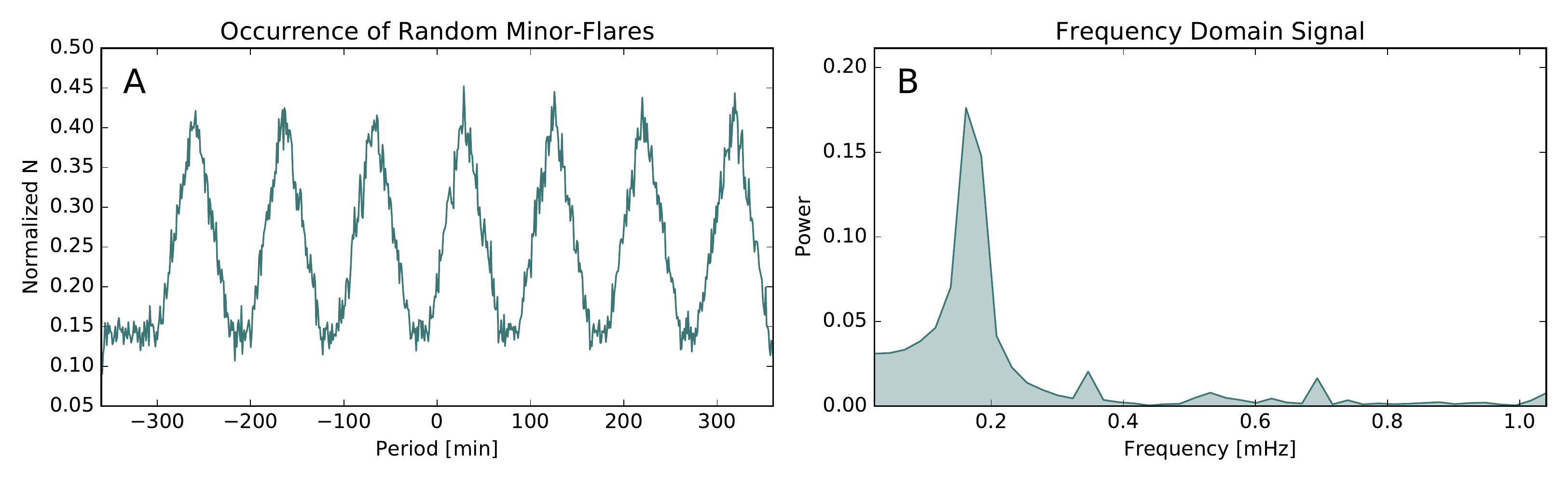}
	\caption{The random-based statistical population for demonstrating the effect of the RHESSI orbital period. The left panel shows the occurrence of the random population and the right panel demonstrates the frequency domain signal.}
	\label{fig2t}
\end{figure*}	

Panel $A$ of Figure \ref{fig1} and Figure \ref{fig2} shows the time variation of the minor flare occurrence based on the RHESSI and GOES flares separately before and after the occurrence of the major event. Both time series show an undesirable feature, namely, the number of the minor-flares is significantly decreased as the occurrence of the major flare is imminent. This behaviour is likely to be the consequence of the enhanced particle emission by the major eruption. The particle emission of a major flare may surpass the fine dynamics of the reconnection in the same active region. The trend can be removed by applying the lag operator $LZ(x_{t}) = Z(x_{t-1})$. The first difference $\Delta Z(x_{t})$ of the RHESSI and GOES populations are defined by, 

\begin{equation}
	\Delta Z(x_{t}) = (1-L)Z(x_{t}).
	\label{sdiff}
\end{equation}

Panel $B$ Figure \ref{fig1} show the temporal variation of the GOES micro-flare Z-Scores after calculating the first difference of the time series. The GOES signal is now suitable for frequency analysis. Unfortunately, the temporal variation of the RHESSI micro-flare occurrence histogram (Panel $A$ of Figure \ref{fig2}) also shows another unwanted feature. The period of this unwanted fluctuation is around 100 minutes. The period may be the consequence of the orbit of the RHESSI satellite, which features 60 minutes observational period and a further 40 minutes black out. 

The orbit itself therefore may influence our sampling method, hence the importance of the dominant 100-minute RHESSI oscillation cannot be certain. For that reason, the following additional analysis is carried on. A virtual satellite is created \textit{in silico}, which observes random eruptive events. The orbit of the virtual satellite features the same properties as the orbit of the RHESSI satellite, i.e. 60 minutes "observational" period and a further 40 minutes black out. The occurrence and the magnitude of the major and minor flares are random, however, the total of random events equals to the number of real observed data. Hence, we modelled 301 major and 4119 minor eruptions. Let us now apply the same methodology for analyses as before with this new random-based sample. Panel $A$ of Figure \ref{fig2t} shows the result of the random-based sample statistics based on 10 thousand simulations. Panel $B$ of Figure \ref{fig2t} demonstrates the frequency domain signal, which (not surprisingly) reveals one dominant peak around 100 minutes (or around $0.17$ mHz). Judging by only the RHESSI dataset, it cannot be safely assumed that this period is only an artefact and there is no other physical process with a similar periodicity. The GOES statistical population, however, does not seem to show similar periodical behaviour. Therefore, the 100-minute oscillation in the RHESSI dataset can be confirmed as an artefact, which can be removed by applying the seasonal lag operator $L^{S}Z(x_{t}) = Z(x_{t-S})$. If the data show fluctuation patterns at every $S$ observations, seasonal difference can be applied for removing the seasonal trend from the time-series by using the expression,

\begin{equation}
	\Delta^{S} Z(x_{t}) = (1-L^{S})Z(x_{t}).
	\label{sdiff}
\end{equation}

Unfortunately, the seasonal lag operator introduces an unwanted consequence. The transformed signal is truncated and the range of signal is shorter than the original. However, since the 100-minute orbital periodicity is so clear, the trend can be extrapolated forward in time before applying the operator, which preserves the time domain to $\pm 360$ minutes. We have chosen a sine function as a model function for describing the orbital period of the RHESSI satellite. The seasonal lag operator is applied on the extended time series. Therefore, the transformed signal remained in the range of $\pm 360$ minutes and the orbital periodicity is removed as well. Panel $B$ of Figure \ref{fig2} shows the temporal properties of the RHESSI minor-flare sample after applying the first differencing and seasonal differencing techniques. The transformed RHESSI and GOES time series are now more suitable for frequency analysis because the results of the temporal analysis are less likely biased by artificial periods due the applied sampling methods.

\section{Frequency Analysis}

We use fast Fourier transform (FFT) for studying the periodic behaviour of the data. The stationarity time sample is required because the FFT algorithm is not able to reveal the local properties of the time-frequency space. By applying the first and seasonal difference methods our time samples fulfil this requirement. Therefore, it is assumed that the periodic behaviour of the signal is time-independent.

The significance level is calculated by employing an Autoregressive Model AR(1), also known as red noise or $1/f$ noise distribution \citep{weedon2003time}. The red noise is a common assumption in astrophysical time series. The power spectrum of the red noise is weighted towards the low frequencies, however, there is no preferred period over the range \citep{kasdin1995discrete}. For estimating the significance of the peaks in the time-frequency space, we generated 1 million independent simulations based on the best-fit AR(1) models. We estimated the coefficients of the applied models. For the GOES statistics, the expression can be written as follows:

\begin{equation} 
	Z(x_t) = 0.2317  Z(x_{t-1}) + \epsilon_t.
\end{equation}

\noindent
In case of the RHESSI statistics, the fitted autoregressive AR(1) expression becomes,

\begin{equation}
	Z(x_t) = 0.2160  Z(x_{t-1}) + \epsilon_t,
\end{equation}

\noindent
where the parameter $Z(x_t)$  is regressed from the previous value $Z(x_{t-1})$ and the parameter $\epsilon_t$ represents the error. The obtained expressions now can be now use for generating simulations. Since the original data values are generally low numbers the generated simulations are based on Poisson distribution rather then Gaussian. Finally, we applied the same methodology than in case of real data, i.e. the simulated data are differenced and FFT is performed as well. The standard deviation and average in each frequency bins are calculated based on the 1 million simulations. In the further statistics, the significance level is defined by the total of the average and three standard deviations.

Panel $C$ of Figure \ref{fig1} shows the result of the period analysis based on the transformed GOES, which is presented by the Panel $B$ of Figure \ref{fig1}. The power spectrum unveils multiple significant frequencies. The lowest frequency is now $1.47$ mHz. The power of the higher frequencies are more pronounced and the peaks appear at  $2.97$, $4.44$, $5.94$ and around $7.43$ mHz. The sampling frequency is $16.65$ mHz and the Nyquist frequency is $8.325$ mHz. Therefore the latest period is close to (but still below) the Nyquist frequency. Other frequencies close the significance level are also visible, such as $7.04$ and $7.79$ mHz. Their significance is still above the significance level $2 \sigma$, however, they are relatively close to another more significant peak. Therefore, these peaks are omitted form the statistics. Note, that the signature of the noise in the power spectrum does not show red noise behaviour. The error is weighted towards the high frequencies, which is a typical blue noise signature. The changed properties of the noise structure is more than likely to be the consequence of the data differentiation.

Panel $C$ of the Figure \ref{fig2} displays the result of the frequency analysis based on the RHESSI flare population.  Artificial periods (100 minutes or 45 minutes fluctuations) due the orbit of the RHESSI satellite and the influence of the South Atlantic Anomaly are removed from the original signal, hence significant peaks at $0.37$ and $0.18$ mHz are not detectable. Unfortunately, the previously detected $1.47$ mHz GOES oscillation is also not detectable in this statistics. However, in the power spectrum of the RHESSI data, there is a significant oscillation around $1.95$ mHz. The other GOES frequencies can be clearly verified by RHESSI observations. Clear and strong $3.15$,  $4.55$, $5.78$ and $7.2$ mHz oscillations are found above the $3\sigma$ threshold. The $5.78$ and $7.61$ mHz oscillations are surrounded by several additional significant peaks, however, these peaks are relatively close to each other.

\begin{figure}
	\centering
	\includegraphics[width=0.97\columnwidth]{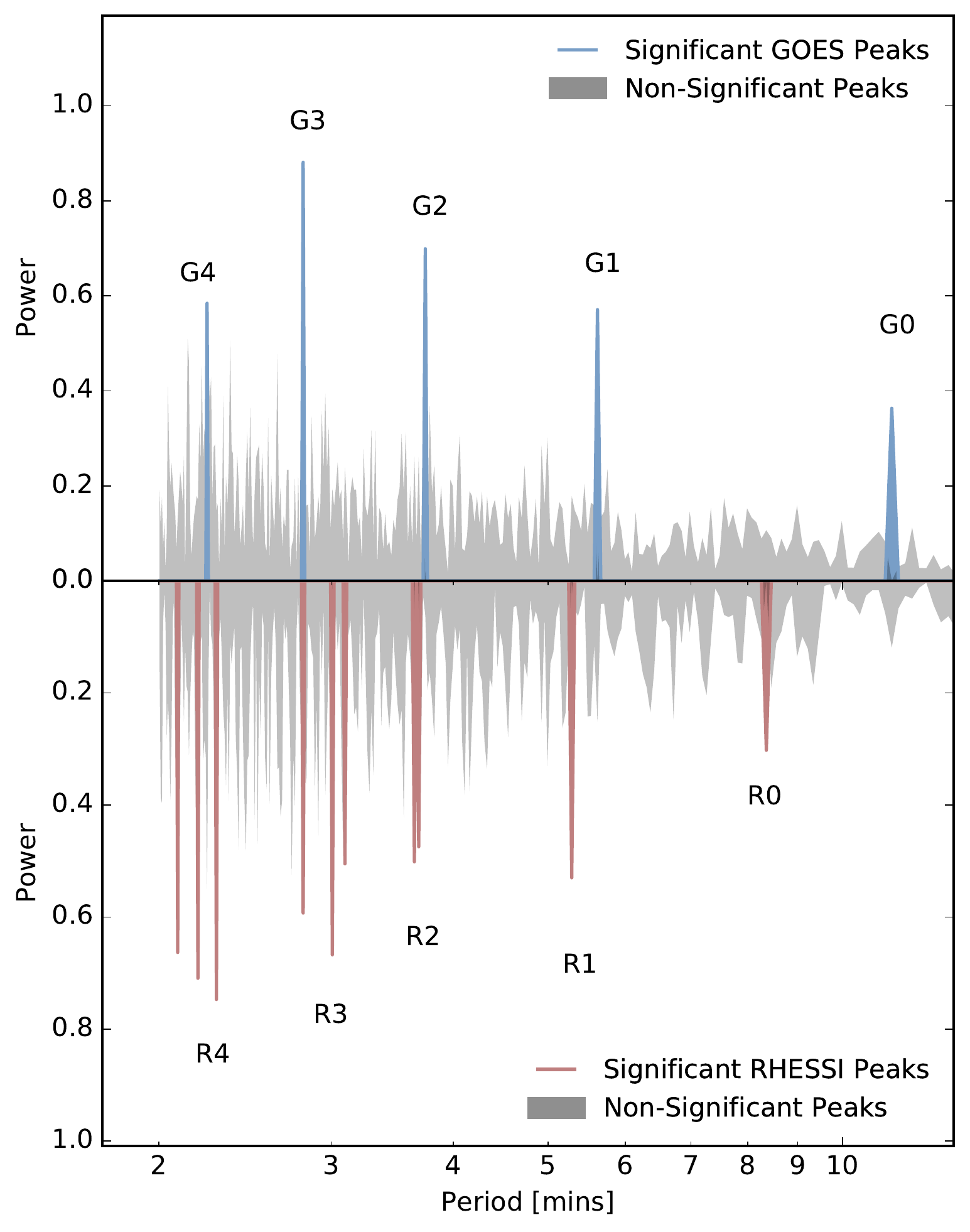}
	\caption{The top panel demonstrates the power spectrum of the GOES statistical sample. The bottom panel also visualises the results of the FFT based on the RHESSI data. The silver peaks are marked as non-significant peaks and the blue and red peaks are greater than the $3 \sigma$ significance threshold.}
	\label{fig3}
\end{figure}

Table \ref{table} and Figure \ref{fig3} display a summary of the obtained peaks for both the GOES and RHESSI statistics. In Figure \ref{fig3}, the GOES statistics show 5 significant oscillation periods, labelled $G0$, $G1$, $G2$, $G3$ and $G4$. The RHESSI statistics displays five remarkable oscillations, labelled  $R0$, $R1$, $R2$, $R3$ and $R4$. The periods $R0$, $R1$ and $R2$ are strong and remarkable oscillations with a single peak structure. However, the remaining two significant oscillations $R3$ and $R4$ each contain three significant peaks. The differences between the first and last peaks in $R3$ and $R4$ are few seconds, hence these peaks cannot be considered with a high confidence to be manifestations of different physical processes. These peaks are considered together and the oscillation period of the $R3$ and $R4$ clusters are calculated by the average of the peaks within. Table \ref{table} shows the average periods of the obtained peaks.

The oscillations $G1$ and $G2$ are clearly confirmed by the peaks $R1$ and $R2$. The $G1$ and $R1$ periods show only $6\%$ difference and the difference between the $G2$ and $R2$ periods is only around $3\%$. The discrepancies of the average periods of the $G3$, $R3$ and $G4$, $R4$ peaks are also negligible (around $3\%$ and $2\%$). However, the discrepancy between longest period $R0$ and period $G0$ is around $24\%$.

\section{Physical Interpretation}

\begin{table}
	\centering
		\caption{Detected Oscillation Frequencies and Periods}
		\label{my-label}
		\begin{tabular}{lllrrr}
			N & Frequency & Period & ID. & Ratio  & Sample \\
			 &  [mHz] &  [min] &  & $ P_{1} / P_{N} $  &  \\		
			\hline
			$1$ & $1.47$ & $11.33$ & G0 & $-$ & GOES\\
			$2$ & $2.97$ & $5.61$ & G1 & $2.01$ & GOES\\
			$3$ & $4.44$ & $3.75$ & G2 & $3.02$ & GOES\\
			$4$ & $5.94$ & $2.80$ & G3 & $4.04$ & GOES\\
			$5$ & $7.43$ & $2.24 $ & G4 & $5.05$ & GOES\\
			\\
			$1$ & $1.95$ & $8.54$ & R0 & $-$ & RHESSI\\	
			$2$ & $3.15$ & $5.28$ & R1 & $1.61$ & RHESSI\\	
			$3$ & $4.55$ & $3.66$ & R2 & $2.33$ & RHESSI\\
			$4$ & $5.78$ & $2.88$ & R3 & $2.96$ & RHESSI\\	
			$5$ & $7.61$ & $2.19$ & R4 & $3.89$ & RHESSI\\	
				\end{tabular}
	\label{table}
\end{table}

Let us now consider a simple oscillatory system: linear transversal waves in a 1-dimensional uniform finite string with length $L$ and fixed endpoints. When this system is perturbed, standing waves will form. The string has a number of specific frequencies, defined by the properties of the string, at which it will naturally vibrate. These frequencies are called eigenfrequencies, where the longest period is called the fundamental mode and the other modes are referred to as higher harmonics. For a uniform string, the fundamental mode $P_{1}$ is described by:

\begin{equation}
	P_{1} = \frac{2L}{c_{ph}},
\end{equation}

\noindent
where, $c_{ph}$ is the phase speed depending on the physical properties of the waveguide. The ratio of the period of the fundamental mode $P_{1}$ to the period of harmonics $P_N$ is:

\begin{equation}
	\frac{P_{1}}{P_{N}}  = N,
	\label{can}
\end{equation}

\noindent
where $N>1$ is an integer for uniform strings, representing the higher harmonic number. For a non-uniform string the period ratio may deviate from its canonical integer value given by Eq. \ref{can} above.

In our study, we propose that the observed oscillation pattern in flare occurrence is driven by the global oscillation of the solar atmosphere, manifested in periodic re-arrangements of the magnetic field. Let us now model the \textit{global}, large-scale solar atmosphere as a simple, uniform and homogeneous 1-dimensional physical environment analogue to the string example above. Here, the basic assumption is that, this solar environment (shaken by a yet to be determined mechanism) responses as a global body to perturbations. Since the solar atmospheric plasma is embedded in magnetic field, field lines will be shaken too, resulting in casual periodic reconnections that are observed as, e.g., RHESSI flares. Here, as a first approximation for insight, we ignore the complexity of active regions, stratification and structuring. All these features influence the period ratio and eigenfunctions of the eigenmodes (in flux tube and solar magneto-seismology context, see e.g. \citealt{2007A&A...462..743E, 2008A&A...486.1015V, 2009SSRv..149....3A, 2012ApJ...748..110L}). Therefore, these omissions may need to bear in mind, and could be potentially important for a deeper diagnostic insight.

Here, we propose that the 3D magnetic solar atmosphere, in an active region, simply responses to some external driver (or drivers) as a resonator. The situation is very similar to that of the solar interior, addressed in great detail by the science of helioseismology. However, here, magnetism is essential. In a sense, our modelling extends and generalises the concept of helioseismology by considering magnetism and applying it to the \textit{upper} solar atmosphere as well. If our proposed thought experiment captures valid physics, and so far the indication of the detected frequencies are mounting evidence towards that, the interpretation of the RHESSI flare observations have the potentials to open up a very new branch of diagnostics branch in solar physics. The key point, why for solar atmospheric diagnostics this may be a major step is, that, the deviation from the canonical values of frequency ratio can be directly linked to obtaining even sub-resolution information about the waveguide \citep{2007A&A...462..743E} associated $P_{1}/P_{N}$ ratio of periods. In the GOES sample, the ratio of the fundamental mode period $P_{1}$ and the period of the first harmonic $P_{2}$ is around the canonical value of $2.01$, indicating that there may not be large-scale strong inhomogeneity.

In general, with the current resolution we cannot yet determine with high confidence how much the true deviation is from this canonical value (referring to a uniform and homogeneous plasma) there is. The observed periods line up as Table \ref{table} shows, the $P_{1} / P_{N}$ ratios follow an order of succeeding integer numbers, if we only consider the GOES statistical sample. If the GOES fundamental period is taken into account as a RHESSI fundamental mode, the canonical values line up similarly as in the GOES sample. If the first significant RHESSI period $R0$ is considered as the fundamental mode, $P_{1} / P_{2} = 1.61$ ratio significantly departures from canonical value of $2$, which may indicate inhomogeneity in the system.

\section{Discussion}

One need more insight into the nature of the underlying oscillations: Are they the result of Alfv\'en, slow or fast standing MHD waves. Is there perhaps coupling between modes. What is the true geometry of the wave guide? Actually, what is the waveguide: is it bounded by the low photosphere and the upper turning point being the chromosphere, or are there other boundaries (reflective or open)? Are the modes linear or nonlinear? If the latter, this is a very difficult mathematical problem to model and proceed with. Answering all these questions indicate the direction of our future aims because they may contribute to understanding the nature of the long-period global oscillations.

Improving our methodology is one of our future aims. Performing a wavelet analysis of the processed x-ray time-series may reveal information about period modulation. The detected oscillations may be present continuously around the major flare source and migrate to different frequencies as time develops. As \cite{2011A&A...525A.112R} concluded, the p-mode leakage upwards along the active region magnetic field lines can play a role in the generation of periodic phenomena, particularly with similar periods to those established in the present work. This needs to be compared more, since the presence of a global driver is more challenging to initiate, while sunspot p-mode waves are ubiquitous. Therefore, developing our physical interpretation and investigating other theories are also future aims.

It is also important to emphasise that the GOES and the RHESSI satellites observe in different wavelengths. It may be possible that discrepancies between the GOES oscillations and the RHESSI oscillations are the consequence of the different observational wavelengths, therefore the different wavelengths may reveal different physical processes.

\section{Conclusion}

We propose that the x-ray flux oscillations are a consequence of the global upper atmospheric oscillations and that, as said before, periodic reconnection is likely triggered by the driving global atmospheric oscillations.

\section*{Acknowledgments}
RE is grateful to STFC (UK), grant number ST/M000826/1, and The Royal Society for support received. This research has made use of SunPy, an open-source and free community-developed solar data analysis package written in Python \citep{2015CS&D....8a4009S}. The authors are indebted to the anonymous reviewer for providing insightful comments and directions for this paper.

\bibliographystyle{aasjournal}
\bibliography{Bibliography}

\end{document}